# Judges matter more than papers in post-publication research assessment

Robert Ward$, Alex Jones§, Lutz Bornmann*

$Bangor University

School of Psychology

Brigantia Building; Penrallt Road

Bangor UK LL57 2AS.

Email: r.ward@bangor.ac.uk

§Swansea University

School of Psychology

Institute of Life Sciences Building 2

Swansea UK SA2 8PP.

Email: alex.l.jones@swansea.ac.uk

*Science Policy and Strategy Department

Administrative Headquarters of the Max Planck Society

Hofgartenstr. 8,

80539 Munich, Germany.

E-Mail: bornmann@gv.mpg.de

Corresponding author.

## Abstract

Research assessment relies on expert evaluations, yet human judgement is noisy, and it is unclear whether differences in assessment arise primarily from differences in genuine research quality or from unwanted differences between evaluators. While numerous studies highlight disagreement and biases in research assessment, they have not quantified judge-related noise relative to variation in the evaluated works. Here we show, in a large post-publication peer review database, that research assessment is driven more by differences between evaluators than by difference in the evaluated research. We partition variance in 239,521 research quality ratings assigned by 12,649 judges to 193,128 papers from the H1 Connect post-publication peer review platform. Using multilevel models, we decomposed judge-related variation into differences in overall severity and differences in the weighting of scientific attributes. We found that judge-related effects accounted for substantially more variance in ratings than the evaluated papers. In our most detailed model, judge-level effects and judge-specific slopes explained 61% of the total variance, whereas combined paper and journal-level effects accounted for only 7%. By contrast, examined measures of directional bias, such as author gender and global affiliation, explained less than 1% of the variance. We conclude that assessment outcomes were shaped more by the judges than by the papers themselves. Our results demonstrate the necessity of noise audits in high-stakes scientific evaluation.

## Introduction

Research assessment depends on expert judgement (Bornmann, 2011). Decisions on publication, funding, hiring, and reputation are based on expert evaluations, meant to reflect underlying research quality. However, human judgement is inherently noisy, and disagreement among experts is found across domains, even when evaluation criteria are well defined (Allison et al, 2014; Bornmann et al., 2011; Cole et al, 1981; Dror & Cole, 2010; Pier et al, 2018; Simonsohn, 2007). Disagreement does not in itself imply system failure. For example, if research differs widely in quality but judges vary only slightly in their assessments, disagreement may have little overall effect on ranking outcomes. The central question is therefore not whether judges disagree, but whether variation in research assessment is driven primarily by differences in the quality of the evaluated works, or by differences between the experts doing the evaluations. The value of research assessment depends on this balance: Research assessment should be based on the quality of the research being evaluated, not on the personal preferences of the judges.

A useful framework for understanding variability in judgement distinguishes bias from noise. Bias refers to systematic deviations in evaluation, such as favouring authors, institutions, or research outcomes. Noise, in contrast, refers to variability in judgements that should in principle be the same. As emphasised by Kahneman et al. (2021), variability that appears as noise at one level of analysis may reflect systematic differences at another. For example, judges might consistently differ in their overall severity (what Kahneman et al., 2021, refer to as "level noise") or in the scientific attributes they value most ("pattern noise"). Once these tendencies are modelled, the apparent disagreement becomes structured variation. The crucial point is that variability between judges can be decomposed into structured components.

The decomposition approach highlights deficits in our understanding of research assessment. Numerous studies have investigated how research evaluation is influenced by factors such as author identity, institutional affiliation, gender, or research outcomes (e.g., Bornmann, 2011; Eyre-Walker & Stoletzki, 2013). Noise in research assessment has received less attention, despite its potential for distorting outcomes (e.g., Boudreau, 2026; Ward, 2026; Bonavia & Marin-Garcia, 2023). Existing studies typically assess disagreement between judges through inter-rater reliability analyses, resubmission designs, or related approaches (Bornmann, 2011; Bornmann et al., 2011; Marsh et al 2008). Although previous studies reveal that disagreement is substantial, they have not, as far as we are aware, decomposed variability into different forms of noise, nor quantified judge-related relative to variation between the evaluated works themselves.

Here we address these challenges using multilevel models that partition variance in ratings of research quality between judges and works. Our approach further allows judge-related variation to be decomposed into systematic components, including differences in overall severity and differences in how attributes of the work are weighted. We apply the approach to a large post-publication peer review dataset from H1 Connect.

## Results

We analysed 239,521 ratings assigned by 12,649 judges to 193,128 papers published across 5,038 journals in the H1 Connect post-publication peer review (PPPR) platform. The platform is a curated system in which experts evaluate published biomedical research using a three-level recommendation scale (*Good*, *Very good*, *Exceptional*), alongside structured classification tags chosen to describe the nature of the paper's contribution (e.g., *Good for* *teaching*). Because evaluations are signed and linked to individual judges, the dataset permits repeated observations of judge behaviour across multiple papers.

## Judge-level variation exceeded paper-level variation

We first quantified how variation in ratings was distributed between papers and judges, using a multilevel model predicting ratings as a continuous variable from random intercepts for papers and judges. Paper-level intercepts capture differences between the evaluated works, while judge-level intercepts capture differences in overall rating between judges ("level noise"). Variation in ratings attributable to judge severity exceeded variation attributable to the papers themselves, 26% to 18% respectively. That is, paper ratings depended more on the people making the ratings than on the papers themselves.

## Journal effects did not explain judge-level variation

We extended the baseline model to include random intercepts for journals, which accounted for 5% of total variance in rankings, and statistically improved model fit, $X^2(1) = 25803$, $p < .0001$. Inspection of journal intercepts showed as expected that journals with the highest positive effects included high-impact biomedical outlets such as *Nature*, *Science*, and *Cell*. Journal effects could potentially operate at either or both judge and paper levels. If judges systematically favoured particular journals, then some of the variance attributed to judges would instead be absorbed by journal effects. However, we found instead that random intercepts reduced variance attributable to papers (from 18% to 10%), but not to judges (from 26% to 25%). This is the expected pattern if journals varied in average paper quality, but judges did not systematically favour particular journals.

## Unmasked pattern noise revealed structured variation between judges

Beyond differences in severity, judges might differ systematically in how they weight different attributes of scientific work. If not modelled explicitly, such differences would appear as what Kahneman et al. (2021) refer to as pattern noise. The classification tags in H1 Connect offer one way to assess different patterns of evaluation between judges. Twelve classifications were available to H1 connect experts (in our dataset) to describe the nature of a paper's contribution and delivery, including novelty, methodological innovation, and conceptual significance. More than 95% of evaluations included at least one classification. Classifications were coded as binary variables indicating whether a particular classification had been assigned by a judge to a given paper. For easier computation and interpretation, we reduced the twelve binary classifications to a smaller set of six latent factors as suggested by parallel analysis, which together explained 65% of variance in classification usage.

We incorporated these latent factors into the previous model by including them as fixed effects and by fitting judge-specific random slopes for each factor. The fixed effects estimated the average relationship between each latent dimension and ratings across all judges, while the random slopes allowed judges to weight these dimensions individually. This model therefore decomposed judge-related variation into differences in overall severity ("level noise") and differences in the weighting of different attributes of scientific work ("pattern noise").

These additions statistically improved model fit, $X^2(12) = 6589$, $p < .0001$. The explanatory power of the model came from its random effects (marginal $R^2 = .018$; condition $R^2 = .42$). Ratings were far more strongly associated with who evaluated a paper than with the paper being evaluated. Variance partitioning showed that judge-level effects (11.5%) plus the sum of judge-slope effects (49.5%) accounted for 61% of total variance in ratings, as compared to 7% for combined paper and journal-level effects. Much of what appeared as unexplained variability in simpler models reflected systematic differences in how judges evaluated the same paper characteristics.

## Findings were robust across model specifications

Across all models, judge-related effects accounted for substantially more variance in ratings than paper-related effects. We therefore conducted a series of robustness checks to assess whether this conclusion depended on particular modelling assumptions or dataset features. Figure 1 summarises our main results and robustness checks.

To assess any disproportionate effect of "singletons", we repeated our analysis on a restricted dataset comprising papers rated by at least two judges who themselves had rated at least two papers. Results from this "no singletons" model closely replicated the original variance structure for paper effects, journal effects, and judge slopes. The main difference was that judge-level variance was reduced. Our original estimates of judge severity may therefore partly reflect sparse information about judges who contributed only a single rating. This difference is also consistent with the possibility that more experienced reviewers may have converged on a set of severity norms. Importantly, the large influence of pattern noise and judge-specific weighting of paper characteristics was little affected, and the general pattern of judge effects outweighing paper effects was stable.

We further assessed robustness of judge slopes by repeating the full model on a restricted dataset comprising cases with at least two classification tags. This ensured the latent factor representations were based on richer paper descriptions. Our results again closely replicated the original variance structure for papers, journals, and judges. This suggests pattern noise effects were robust and not driven by sparsely classified papers.

In contrast, permutation testing, in which latent-factor assignments were randomly shuffled, reduced variance attributable to judge-specific slopes from 49% to 0.6%. The collapse of judge-slope variance under permutation is expected if the pattern-noise effects in our full model reflect meaningful structure in judge evaluations, rather than artefacts of model flexibility or overfitting.

Finally, we took an entirely different approach using a Bayesian model that assumed an ordinal rather than continuous rating scale. We refit the baseline model, estimating variance accounted for by judges, papers, and journals, and found a very similar outcome, indicating that our results did not depend on treating ratings as continuous, and demonstrating that different modelling approaches led to similar conclusions.

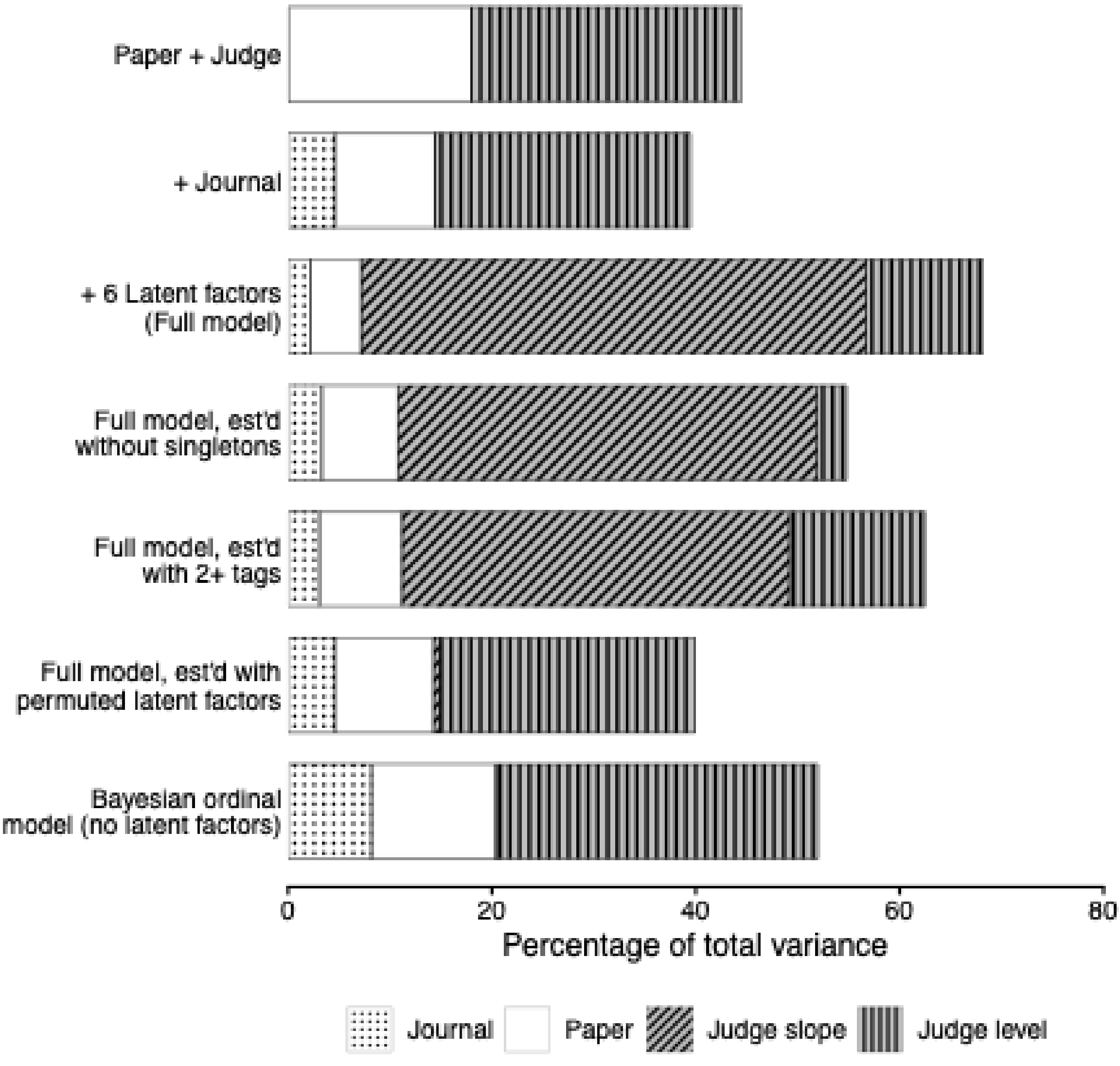

Figure 1. Variance partitioning across primary models and robustness analyses. Each horizontal stacked bar shows the proportion of variance in ratings attributable to papers, journals, judge-level intercepts, judge-level slopes, and residual variation. The first three models progressively develop the full model: paper and judge intercepts; addition of journal effects; and addition of judge-specific slopes on six latent paper dimensions. These models show a consistent pattern in which variance attributable to judges exceeds that to papers. The remaining models provide robustness checks. Restricting analyses to non-singleton observations and to reviews containing richer classification information produced partitions similar to the full model. In contrast, when latent-factor scores were randomly permuted across cases, variance attributable to judge slopes collapsed to near zero, indicating that the slope effects depend on meaningful relationships between paper characteristics and ratings rather than model structure alone. Finally, a Bayesian ordinal model based on paper and judge intercepts closely replicated the variance partition obtained from the linear mixed models. Overall, the variance structure was highly stable across specifications, with judge-level effects consistently accounting for more variance in ratings than paper-level effects.

### Measures of bias explained little variance

Although our primary concern is understanding noise, we examined several potential sources of directional bias. We extended the baseline model to include gender and global affiliations of authors and experts (see the Appendix for details). Although some effects were statistically detectable in the large dataset, none explained a practically important proportion of variance (< 1%) or changed the balance between variance explained by judges versus papers. The problem is noise, not bias.

## Discussion

Previous peer review research has examined the extent to which reviewers agree or disagree in their assessments of a manuscript. Our central motivation was not whether judges disagree, but whether variation between judges is large or small relative to variation in the quality of the work being evaluated. Across all our models, judges accounted for more variation in quality than the papers themselves. Differences were systematic at the level of individual judges but heterogeneous across judges. This finding was robust to alternative model specifications, different assumptions about the rating scale, and restricted datasets. As we introduced richer representations of paper characteristics into our models, we found an increasing proportion of what initially appeared as residual noise became attributable to judge-specific evaluation patterns. In our most detailed model, judges explained more than seven times the variation than did papers (60% to 8%).

Our results carry significant implications for research assessment. Studies of peer review frequently focus on measures of agreement between reviewers, and this focus clearly supplies valuable information. However, we argue that our approach, simultaneously modelling variability of papers and judges, is important for evaluating judge variability in context. This may be best thought of as a “noise audit” partitioning variation between the judges and the judged and identifying the sources of judge variability. Based on our findings, assessment outcomes were shaped more by the judges than the papers themselves.

Our results indicate that discussions of research assessment may benefit from distinguishing more clearly between bias and noise. Although concerns about bias frequently dominate debates about peer review, the measured effects of gender and global affiliation in this study were small and explained little variance in ratings. This is encouraging evidence that post-publication peer review is relatively free from demographic and affiliation-related biases, at least within the H1 Connect system. By contrast, judge-related noise effects were large and frequently reflected systematic differences in evaluative weighting. This does not imply that bias is absent, only that in the present dataset variability attributable to evaluator-related noise was considerably larger than variability attributable to the specific bias-related factors we examined. Although bias has been a major focus of peer reviewed research to date, our findings suggest noise may be the more serious issue: The problem is noise, not bias.

Although H1 Connect has the virtue of being an exceptionally large and well-curated biomedical research database, the database differs from some forms of scientific evaluation. Peer review for journal articles, like final decisions on grant applications, is more concerned with a binary yes/no decision than a metric rating as in H1 Connect. That is, H1 Connect is a recommendation rather than gatekeeping system. The relative contributions of judge- and paper-related variation may differ in these contexts, although the extent of any differences is unknown. However, there are similarities between H1 Connect and other research evaluation systems, including high-stakes exercises such as the UK Research Excellence Framework (REF). Both H1 Connect and the REF evaluate research after publication using a limited rating scale. In each case, the work being assessed has already passed an initial quality threshold. In the REF, this threshold is typically established through departmental review and internal selection processes. In H1 Connect, it arises through experts' decisions about which papers to evaluate. In both systems, the primary challenge is not identifying clearly flawed research but distinguishing between different levels of good research. This is precisely the setting in which evaluator-related variation may have its greatest consequences, where a compressed range of quality is assessed through a noisy evaluation process.

Kahneman et al (2021) argued wherever there is human judgement there is noise. Our findings suggest research assessment is no exception. The critical issue however is not that noise exists, but its magnitude relative to variation in research quality. Ward's (2026) modelling of the REF demonstrated that even moderate levels of judge noise can propagate into substantial distortions, a systematic bias against the best departments, and highly unstable department rankings. The present findings complement this work by providing direct evidence from a real-world evaluation system that judge-related variation can exceed variation between the research outputs. Together these findings suggest judge-related noise is not simply an unavoidable feature of research assessment, but a measureable property of evaluation systems. We therefore suggest that routine noise audits, paritioning variance between judges and research outputs, should be a standard component of high-stakes scientific evaluation.

# Method

## The H1 Connect dataset

Data were obtained from H1 Connect (https://connect.h1.co), a post-publication peer review platform built on the long-standing F1000 model. H1 Connect operates an expert-curated recommendation system in which invited specialists evaluate published biomedical and life-science research. H1 Connect provided us with a snapshot from its database, current as of November 2023. Each case in the snapshot consists of an evaluation (rating, text, and classifications) linked to an identified judge and publication. Judges have guidance for the possible ratings of *Exceptional*, *Very Good*, and *Good*. *Exceptional* (three stars) is the highest level of evaluation and is reserved for field defining, paradigm shifting, or methodologically groundbreaking work. *Very good* (two stars) indicates a substantial and noteworthy contribution to the field, offering significant advances, robust evidence, or important methodological improvements. *Good* indicates that a publication is solid, reliable, and of clear interest to specialists, even if its contribution is more incremental. All evaluations are signed, public, and citable. The dataset additionally includes publication metadata, journal information, classification tags, publication and evaluation date, and author affiliations. At the time of analysis, the full dataset consisted of 246,245 cases, of which 4833 were excluded as incomplete.

Available classifications that were used by the experts in our dataset comprised: *New Finding*; *Technical Advance*; *Novel Drug Target*, *Good for Teaching*; *Controversial*; *Interesting Hypothesis*; *Refutation*; *Confirmation*; *Systematic Review*; *Negative Finding*; *Review*. Classifications were coded as binary variables indicating whether a particular classification had been assigned by an expert to a given paper. These tags function as metadata that complements the rating system: while ratings express overall importance, tags articulate specific dimensions of value. Although not required, 95% of evaluations used at least one classification tag.

We did not analyse the binary classifications directly, but a reduced set of latent factors identified by factor analysis on the classifications. Parallel analysis supported a six-factor solution, accounting for 65% of the variance in classification use. Latent factor scores were calculated for each evaluated paper and used as predictors in subsequent multilevel models.

## Multilevel models

Variance in ratings was analysed using linear mixed-effects models implemented in the R package lme4 and lmerTest. Models were estimated using restricted maximum likelihood. The specifications of our primary models below use R formula syntax.

The baseline model included random intercepts for papers and judges:

```
rating ~ (1|paper) + (1|judge)
```

Paper intercepts capture variation attributable to evaluated papers, whereas judge intercepts capture systematic differences in overall rating severity between judges.

Subsequent models incorporated journals as an additional random effect:

```
rating ~ (1|paper) + (1|journal) + (1|judge)
```

To examine judge-specific evaluation patterns, latent factor scores derived from classification tags were included as fixed effects and as random slopes varying by judge. These models allowed the relationship between paper characteristics and ratings to differ across judges:

```
rating ~ Latent_Factor1+Latent_Factor2+…Latent_Factorn +
(1|paper) + (1|journal) +
(0 + Latent_Factor1+Latent_Factor2+…Latent_Factorn|| judge) +
(1|judge)
```

This specification allows the estimated random intercepts for judges to be uncorrelated to the estimated random slopes by latent factors derived from classification tags. This means that the severity of a judge is uncorrelated with their weighting of different latent factors, for example, both lenient and severe judges might weight different factors in the same way.

## Variance partitioning

Variance components were extracted from fitted models and expressed as percentages of total model variance. Judge-related variance comprised random intercept variance associated with judges together with variance attributable to judge-specific slopes where applicable. Paper-related variance was estimated from paper random intercepts. Journal-level variance was estimated from journal random intercepts. Because the inclusion of journals reduced variance attributable to papers but not judges, reported totals for paper-related variance include journal-level effects. This is conservative with respect to our finding that judge effects exceed paper effects.

## Robustness analyses

Additional analyses verified the robustness of our results. The fit to the full model with six latent factors was repeated on a restricted “no singletons” dataset, comprising the 76,698 cases where a paper was evaluated by at least two judges who themselves evaluated at least two papers.

First, all primary models were repeated on a restricted "no singletons" dataset containing the 76,398 cases where the evaluated paper was evaluated by at least two judges who themselves evaluated at least two papers. Second, factor-based models were assessed for sensitivity to sparse classification codes by repeating the analysis using only the 101,584 cases with at least two assigned classifications. Finally, models incorporating six latent factors were estimated using multiple optimisers, comprising nloptwrap, bobyqa, and optimx.

To assess whether judge-specific slope variance reflected meaningful structure rather than model flexibility, permutation analyses were conducted by randomly shuffling the latent factor assignments on a row-wise basis. If for example, individual classifications were shuffled, unrealistic combinations could occur (e.g., *Controversial* and *Confirmatory*). Our approach ensured that the latent factors reflected a realistic structure, even while we disrupted the specific associations of factors to papers. After shuffling, the factor-based multilevel models were then re-estimated on the permuted data.

To assess sensitivity to the treatment of ratings as continuous, baseline models were re-estimated using a Bayesian cumulative ordinal framework. Here, ratings were no longer rated as a continuous variable but were instead cast as manifested scores of a latent unobserved variable, with two cutpoints separating a *Good* and *Very Good* rating, and *Very Good* and *Exceptional*, respectively. The cutpoints, placed along a continuous logistic distribution, allow for a probabilistic generation of each rating. This model formulation complemented the initial judge, paper, and journal structure:

```
rating ~ (1|paper) + (1|journal) + (1|judge)
```

It allowed us to calculate the variation explained in the latent variable attributable to each source with a more realistic modelling of the ordinal, discrete rating scale.

## Bias-related analyses

We conducted a series of exploratory analyses examining potential sources of directional bias in evaluations. These analyses focused on judge and author characteristics rather than the content of the evaluated papers. Potential biasing factors were included as fixed effects and their magnitude assessed by the partial $R^2$ of the fitted multilevel model.

Gender was inferred from names using the Python package gender_guesser, which classifies names based on historical naming patterns. Because these classifications are derived from names rather than self-identification, they should be interpreted as measures of perceived gender rather than actual gender identity. Cases were excluded where gender_guesser returned an unknown or androgynous result. Analysis examined the effects of author gender, expert gender, and author–expert match as fixed effects in the baseline multilevel model. Because papers typically include multiple authors, author gender was operationalised in separate analyses based on first author gender, author gender, and the plurality gender among all listed authors.

Institutional affiliation was coded from author and judge countries. Analyses examined the effects of author country, expert country, and author–expert match as fixed effects. To investigate broader geopolitical patterns, countries were additionally classified according to their location in the Global North or Global South. As with gender, affiliation-based analyses were conducted using alternative author definitions based on first author affiliation, corresponding author affiliation, and the plurality affiliation among all authors.

## References

Allison, K. H., Reisch, L. M., Carney, P. A., Weaver, D. L., Schnitt, S. J., O'Malley, F. P., & Elmore, J. G. (2014). Understanding diagnostic variability in breast pathology: Lessons learned from an expert consensus review panel. *Histopathology*, 65(2), 240–251.

Bonavia, T., & Marin-Garcia, J. A. (2023). A noise audit of the peer review of a scientific article: A WPOM journal case study. *WPOM-Working Papers on Operations Management*, 14(2), 137-166. https://doi.org/10.4995/wpom.19631

Bornmann, L. (2011). Scientific peer review. *Annual Review of Information Science and Technology*, *45*, 199-245. https://doi.org/10.1002/aris.2011.1440450112

Bornmann, L., & Williams, R. (2020). An evaluation of percentile measures of citation impact, and a proposal for making them better. *Scientometrics*, *124*, 1457–1478. https://doi.org/10.1007/s11192-020-03512-7

Bornmann, L., Mutz, R., & Daniel, H.-D. (2011). A reliability-generalization study of journal peer reviews: A multilevel meta-analysis of inter-rater reliability and its determinants. *PLOS ONE*, 5(12), e14331.

Boudreau, K. (2026). Field experiments in the science of science: Lessons from peer review & the evaluation of new knowledge. Retrieved January 26, 2026, from http://dx.doi.org/10.2139/ssrn.6025114

Ceci, S. J., Kahn, S., & Williams, W. M. (2023). Exploring gender bias in six key domains of academic science: An adversarial collaboration. *Psychological Science in the Public Interest*, *24*(1), 15-73. https://doi.org/10.1177/15291006231163179

Cole, S., Cole, J. R., & Simon, G. A. (1981). Chance and consensus in peer review. *Science*, 214(4523), 881–886.

Dror, I. E., & Cole, S. A. (2010). The vision in "blind" justice: Expert perception, judgment, and visual cognition in forensic pattern recognition. *Psychonomic Bulletin & Review*, 17(2), 161–167.

Eyre-Walker, A., & Stoletzki, N. (2013). The assessment of science: The relative merits of post-publication review, the impact factor, and the number of citations. *PLOS Biology*, *11*(10), e1001675. https://doi.org/10.1371/journal.pbio.1001675

Haunschild, R., & Bornmann, L. (2023). Identification of potential young talented individuals in the natural and life sciences: A bibliometric approach. *Journal of Informetrics*, *17*(3), 101394. https://doi.org/10.1016/j.joi.2023.101394

Kahneman, D., Sibony, O., & Sunstein, C. R. (2021). *Noise: A flaw in human judgment*. Little, Brown Spark.

Lee, K. P., Boyd, E. A., Holroyd-Leduc, J. M., Bacchetti, P., & Bero, L. A. (2006). Predictors of publication: Characteristics of submitted manuscripts associated with acceptance at major biomedical journals. *Medical Journal of Australia*, *184*(12), 621-626.

Marsh, H. W., Jayasinghe, U. W., & Bond, N. W. (2008). Improving the peer-review process for grant applications: reliability, validity, bias, and generalizability. *American Psychologist*, 63(3), 160.

Murray, D., Siler, K., Larivière, V., Chan, W. M., Collings, A. M., Raymond, J., & Sugimoto, C. R. (2019). *Author-reviewer homophily in peer review*. Retrieved January 23, 2026, from https://www.biorxiv.org/content/biorxiv/early/2019/08/04/400515.full.pdf

Pier, E. L., Brauer, M., Filut, A., Kaatz, A., Raclaw, J., Nathan, M. J., & Carnes, M. (2018). Low agreement among reviewers evaluating the same NIH grant applications. *Proceedings of the National Academy of Sciences*, 115(12), 2952–2957.

Simonsohn, U. (2007). Clouds make nerds look good: field evidence of the impact of incidental factors on decision making. *Journal of Behavioral Decision Making*. 20 (2): 143–152. https://doi.org/10.1002/bdm.545

# Appendix: Measures of bias explained little variance

To complement the variance-partitioning analyses of judge noise, we examined several potential sources of systematic bias in H1 Connect ratings (Bornmann, 2011). We use the term bias to refer to systematic deviations in ratings associated with characteristics that are irrelevant to intrinsic scientific quality. We specifically assessed influence of judge and author gender (Ceci et al., 2023), judge-author gender match, country of affiliation (Lee et al., 2006), global North–South affiliation (Murray et al., 2019), and judge–author affiliation match. We also examined journal prestige as a potential influence on ratings and whether judges differed systematically in the weight they assigned to prestige.

## Gender bias

Author and judges gender was inferred from first names using the Jörg Michael name database, accessed through the Python package `gender_guesser`. Our analyses are therefore best understood as explorations of bias from perceived gender based on name. Outputs for individual authors and judges were scored as -2 (*male*), -1 (*mostly male*), 0 (*unknown* or *androgynous*), +1 (*mostly female*), or +2 (*female*). Coding author gender for a paper is complicated by the fact that papers typically have multiple authors, who may differ in gender and scientific contribution. We therefore considered several alternative representations. Three paper-level measures were constructed: first-author gender, corresponding-author gender, and the mean gender score across all authors. In addition, as these measures were only moderately correlated ($0.38 < r < 0.51$), we constructed a composite score that summed the first-author, corresponding-author and mean-author measures. For each of these four measures, fixed effects comprised a fixed intercept, author gender, judge gender, and their interaction. Random effects comprised paper, judge, and journal random intercepts.

Two other models shared a similar structure but used categorical coding of authors and judges. Names scored as *male* or *mostly male* were categorised as male, and names scored as *female* or *mostly female* as female. For these two models, papers with unknown or ambiguous names were dropped, resulting in a set of 137k cases.

Results for all models were very similar, regardless of the specifics of gender coding. Marginal $R^2$ in all cases was less than .001. Across all models, the maximum possible change in predicted rating attributable to gender effects was 0.07 points on the 1–3 rating scale. Interaction terms representing the match between author and judge gender were not significant. The variance partitioning related to judge, paper, and journal was not materially affected by the inclusion of these fixed effects. Overall, we found no substantive contribution from gender bias effects in explaining differences in peer-review scores.

## Country affiliation and in-group bias

We assessed potential differences between judges by country of affiliation. Judges were assigned a country of affiliation based on the institutional affiliation provided in the H1 Connect database. Each paper was assigned a country of affiliation based on the plurality of all author affiliations for the paper. We then encoded whether the affiliations of judge and paper matched. From these variables our model took the form:

```
rating ~ match + (1 | paper) + (1 | judge_country) + (match |
judge_country) + (1 | judge_country:judge)
```

This formula models several forms of potential bias. First, the fixed effect of `match` assesses whether there is an overall in-group bias, such that judges rate papers from their own country more or less harshly than those from other countries. Second, the random intercepts for `judge_country` can capture systematic differences in severity based on country (e.g., judges from country A are harsh, from country B are lenient). Third, the random slopes represented by `match|judge_country` allow countries to differ in the strength and even direction of any overall in-group bias (e.g., judges from country A show strong in-group preference, judges from country B show weak out-group preference). The final judge term is not related to bias but provides judges with an intercept relative to their country, `(1 | judge_country:judge)`, allowing individual variation in severity as in all our other models. The fixed effect of match was statistically significant, but the marginal $R^2$ for the model was only .002, and a match provided only an additional .06 rating points on a 1-3 scale. The random intercepts and slopes for judge country explained only 1.1% of variance. In contrast, judge-level intercepts explained 26% of variance, and paper intercepts 18%, just as in our original model.
These analyses indicate only the most limited evidence for national affiliation bias. While a statistically reliable tendency toward in-group favouritism was detectable, its magnitude was of no practical significance.

## Regional affiliation

We also looked for bias related to broader geopolitical affiliation, by classifying both judges and papers according to Global North and Global South (see https://en.wikipedia.org/wiki/Global_North_and_Global_South). Judge affiliation was assigned based on their reported institutional affiliation. Paper affiliation was based on the plurality of author affiliations.
Unlike the previous country-based analysis, with only two regions, the affiliation of judge was better encoded as a fixed effect:

```
rating ~ match + judge_region + match:judge_region + (1 | judge) +
(1 | paper)
```

As before, the fixed effect of `match` tests for an overall in-group bias, asking whether judges systematically rate papers from their own region differently. The fixed effect or judge_region assesses for differences in overall judge severity between regions, and the interaction of judge region by match allows the effect of match to vary between regions. We again include random effects of paper and judge for comparison to our other models.
Judges based in the Global South were predicted to slightly higher ratings (.06 on the 1-3 scale) than judges based in the Global North, $t(12420)=6.97$, $p<.0001$. Ratings were also marginally higher when judge and paper belonged to the same regional grouping (.05 on the 1-3 scale), $t(205400)=12.1$, $p<.0001$ and this matching effect was .03 ratings point larger among Northern than Southern judges, $t(224000)=4.7$, $p<.0001$. The large sample size means all differences are statistically significant; however, marginal $R^2$ for this model was only .001. Again, judge-level intercepts accounted for approximately 26% of the variance, compared with 18% attributable to papers, just as in our original model and the country-based model above.
Regional affiliation therefore produced effects that were statistically detectable, given our large sample, but of negligible effect size, and with no material effect on the overall variance structure of the data.

## Bias associated with journal characteristics

In our *Paper + Judge + Journal* model, journal intercepts might broadly reflect journal prestige. To assess differences between judges in how they weight prestige, we fit an additional model using, as a fixed effect, a citation-percentile measure of journal prestige, centred and scaled to improve convergence. The citation-percentile measure of journal prestige is the median of the citation impact percentile of the papers published in a journal (Bornmann & Williams, 2020). The indicator is a field- and time-normalized metric of journal prestige (Haunschild & Bornmann, 2023). We additionally included judge-specific random slopes for prestige. We reasoned that if judges differed in how journal prestige affected their ratings, then some of the level difference between judges might be reduced with the inclusion of this random slope.

These additions produced a statistically significant improvement in fit over our original *Paper + Judge + Journal* model, $X^2(2) = 781$, $p<.0001$, but with little practical effect. Marginal $R^2$ was .006. Judge slopes for prestige accounted for less than 1% of total variance, and the variance partitioning of paper, judge, and journal was little affected. Paper and journal-level effects together accounted for 14% of variance, while judge-level accounted for 25%. These values are almost identical to our original *Paper + Judge + Journal* model, 15% for paper and journal-level, 25% for judge-level. Therefore, while papers published in more prestigious journals tended to receive slightly higher ratings, individual judges showed little disagreement in the importance they assigned to journal prestige.

## Summary

We assessed several potential forms of systematic judge bias in the H1 Connect dataset. Although several effects reached conventional levels of statistical significance because of the exceptional size of the dataset, effect sizes were always negligible. This is encouraging evidence that post-publication peer review is relatively free from demographic and affiliation-related biases, at least within the H1 Connect system.

However, these findings put the principal results of our study into sharper perspective. The small contribution of the bias variables contrasts with the concerning contribution of judge-level effects found throughout our variance-partitioning analyses. While concerns about systematic bias are well founded and warrant continued attention, the dominant source of variability in research assessment appears to be differences between judges.